\documentclass[aps,prd,twocolumn,superscriptaddress,amsmath,amssymb]{revtex4}
\pdfoutput=1
\usepackage{graphicx}
\usepackage{color}
\usepackage{ulem}
\usepackage{hyperref}
\usepackage{float}
\usepackage{cancel}
\usepackage{pifont}
\usepackage{braket}
\usepackage{bbold}
\usepackage{subfigure} % subfiguras

\begin{document}
	\title{Uncovering Majorana nature through a precision measurement of $CP$ phase}
	
	\author{J.C. Carrasco-Mart\'inez}
	\affiliation{Secci\'on F\'isica, Departamento de Ciencias, Pontificia Universidad Cat\'olica del Per\'u, Apartado 1761, Lima, Per\'u}	
	\affiliation{Department of Physics, University of California, Berkeley, CA 94720, USA}
	\author{F.N. D\'iaz}
	\affiliation{Secci\'on F\'isica, Departamento de Ciencias, Pontificia Universidad Cat\'olica del Per\'u, Apartado 1761, Lima, Per\'u}	
	\author{A.M. Gago}
	\affiliation{Secci\'on F\'isica, Departamento de Ciencias, Pontificia Universidad Cat\'olica del Per\'u, Apartado 1761, Lima, Per\'u}		
	
	\begin{abstract}
We show the possibility to discover 
the neutrino nature by measuring the Majorana CP phase at the DUNE experiment. This phase is turned on by a decoherence environment, possibly originated by physics at the Planck scale. A sizable distortion in the measurement of the Dirac CP violation phase $\delta_{\mathrm{CP}}$ is observed at DUNE when compared with T2HK measurement due to decoherence and non-null Majorana phase. Being that, when the measurement of the Majorana phase is performed at DUNE, it reaches a precision of 23 (21) $\%$ for a  decoherence parameter $\Gamma=4.5(5.5)\times 10^{-24} \mathrm{GeV}$ and a Majorana phase equal to $1.5 \pi$. The latter precision is similar to the one obtained at the T2K experiment at its current Dirac CP violation phase measurement. 

	\end{abstract}
	\maketitle
%\paragraph{Introduction.---}	

\section{Introduction}
   
The origin of neutrino masses is one of the most relevant questions of modern elementary particle physics \cite{Murayama:2006qb,Valle:2015mma}. The Standard Model (SM) Higgs mechanism could generate the neutrino masses if they were Dirac particles. However, this SM alternative does not explain why the neutrino masses are less than one-millionth of the electron's mass, the smallest charged lepton. The general belief is that the latter is resolved through the seesaw mechanism, being the most inexpensive case when neutrinos are Majorana particles \cite{Yanagida:1979as,GellMann:1980vs,Glashow:1979nm,Minkowski:1977sc,Mohapatra:1979ia,Magg:1980ut,Schechter:1980gr,Mohapatra:1980yp,Foot:1988aq,Shrock:1980ct,Konetschny:1977bn,Cheng:1980qt,Lazarides:1980nt}. The Majorana neutrinos, a fermion that is its own antiparticles, imply the total lepton charge violation, a conserved number within the SM processes \cite{Atre:2009rg,Zee:1985id}. Therefore, the quest for elucidating the Majorana nature of neutrinos is one of our best chances to learn about what is beyond the SM.
The usual way to look for Majorana neutrinos is using the neutrinoless double beta decay, a lepton charge violating process not allowed by the SM \cite{Faessler:1999zg,Avignone:2007fu,Vergados:2012xy,Bilenky:2012qi}. Until now, there is no signal from the latter \cite{Agostini:2019hzm,Anton:2019wmi,Adams:2019jhp,KamLAND-Zen:2016pfg} or other proposes way of searches for Majorana neutrinos \cite{Sirunyan:2018xiv,Aaij:2014aba,Balantekin:2018ukw,Bora:2016ygl}.

The standard oscillation neutrino (SO) probabilities take the same form regardless of the neutrino nature. Only the Dirac CP phase is observable while the two other CP violation phases, when the neutrino is Majorana, are absorbed \cite{Giunti:2010ec}. However, if we suppose a new physics(NP) phenomenon, subleading to the SO mechanism, exists, the latter situation could be no longer valid. Therefore, this NP could help the Majorana phases to emerge in the oscillation probabilities.
The NP phenomena we focus on forecasts an interaction between the neutrino system with the environment at the Planck scale level, which is characterized by a foamy space-time \cite{Benatti:2000ph}. The loss of quantum coherence (decoherence) in the neutrino system is the interaction's main signature. The foamy space-time is predicted in the context of strings and branes~\cite{Ellis:1992eh,Ellis:1992pm,Benatti:1998vu}, and quantum gravity~\cite{Hawking:1982dj}. This quantum decoherence phenomena in the neutrino system have been vastly studied in the literature \cite{Lisi:2000zt,Barenboim:2006xt,Bakhti:2015dca,Carpio:2017nui,Carpio:2018gum, Gomes:2020muc} wherein most of the cases containing only the damping effects. Nevertheless, the phenomenology of the interaction between neutrino and a quantum decoherence environment goes beyond the damping effects, being that this might add new contributions to the violation of CP or \textit{CPT} \cite{deOliveira:2013dia}, \cite{Capolupo:2018hrp} and \cite{Carrasco:2018sca}. Through these kinds of contributions in the neutrino oscillation formula, we can reveal the neutrino's Majorana nature. The observability of the Majorana nature using neutrino oscillation is a rather novel approach. 

At this point it is worth mentioning that our work's cornerstone hypothesis is that the Dirac CP violation phase measured at T2K experiment represents its real value since it is unaffected by quantum decoherence in neutrino oscillations. Thus, considering that the effects of a quantum decoherence environment, coupled with a non zero Majorana phase, are sizable in the DUNE data, our strategy is divided into two steps. The first step is to assess how much a measurement at DUNE experiment \cite{Acciarri:2015uup} of the Dirac CP violation phase can deviate from the true one. For this purpose, and taking the pure SO
as a theoretical hypothesis in the fitting, the CP violation phase obtained at DUNE is compared with the projected one at the T2HK experiment \cite{Abe:2018uyc}. The T2HK projected measurement comes to be an upgrade in the precision of the CP violation measurement at T2K. The second step, and final goal in this letter, is to go beyond testing the DUNE accuracy for determining a Majorana CP phase simultaneously with the decoherence parameter.

\section{General Theoretical Formalism}
The treatment for a neutrino subsystem embodied in an infinity unknown reservoir or environment, with which the former interacts weakly,  can be obtained by the Lindblad Master equation~\cite{Benatti:2000ph}:	
	\begin{equation}
	\label{Lindblad}
	\frac{\partial\rho(t)}{\partial t}=-i[H,\rho(t)]+\mathcal{D}[\rho(t)],
	\end{equation}
where $\rho(t)$ is the reduced (neutrino) density matrix, obtained after trace over the degrees of freedom of the environment, $H$ is the Hamiltonian of the neutrino subsystem and $\mathcal{D}[\rho(t)]$ is the dissipative term which encloses the decoherence phenomena. The aforementioned factor can be written as $\mathcal{D}[\rho(t)]=\frac{1}{2}\sum_{j} \left([A_j,\rho(t) A_j^{\dagger}]+[A_j\rho(t), A_j^{\dagger}] \right)$. Thus, if we work in a three-level system the operators 
$\rho$, $H$ and $A_j$ can be written as follows: $\rho=\sum \rho_\mu t_\mu$, $H=\sum h_\mu t_\mu$, and $A_j=\sum a^j_\mu t_\mu$
where $\mu$ is running from 0 to 8, $t_0$ is the identity matrix and $t_k$    the Gell-Mann matrices $(k=1,...,8)$ that satisfy $[t_a,t_b]=i\sum_c f_{abc}t_c$, where $f_{abc}$ are the structure constants of $SU(3)$. The 
hermiticity of $\hat{A}_j$, which is secured demanding a time-increase 
Von Neumman entropy, allow us to write a symmetric  $\bf{D}\equiv$ $ D_{kj}$ dissipative/decoherence matrix as: $D_{kj}=\frac{1}{2}\sum_{l,m,n}(a_{nl})f_{knm}f_{mlj}$, where $a_{nl}=\vec{a}_n.\vec{a}_l$ with components $D_{\mu 0}=D_{0 \mu}=0$, and $\vec{a}_{r}=\{a^1_r, a^2_r, ...,a^8_r \}$. The matrix $\bf{A}\equiv$ $ a_{nl}$ should be positive~\cite{Benatti:2000ph} in order to fulfill the complete positivity condition which states that 
the eigenvalues of the mixing matrix $\rho(t)$ should be positive at any time. Besides, given their inner product structure, the $D_{kj}$ must satisfy the Cauchy-Schwartz inequalities. Adding the conservation of the probability to the aforementioned conditions we get 
the following evolution equation for $\rho(t)$:
	\begin{equation}
	\label{Evolutionequation}
	\dot{\rho}_0=0,\hspace{0.5cm}\dot{\rho}_k=(H_{kj}+ D_{kj}) \rho_j =M_{kj}\rho_j,
	\end{equation}
	where $H_{kj}=\sum_i h_i f_{ijk}$. The matrix form of the solution of Eq.~(\ref{Evolutionequation}) is:
	\begin{equation}
	\label{SolEvoleq}
	\varrho (t)=e^{{\bf M}t}\varrho (0),
	\end{equation}
where $\varrho$ is an eight column vector compose by the $\rho_k$ and $\bf M$$\equiv M_{kj}$. Hence, the neutrino oscillation probability
$\nu_\alpha\rightarrow\nu_\beta$ is given by:
	\begin{equation}
	\label{GeneralProbability}
	P_{\nu_\alpha\rightarrow\nu_\beta}=\frac{1}{3}+\frac{1}{2}(\varrho^{\beta}(0))^T\varrho^{\alpha} (t),
	\end{equation}	
or written in terms 
of the coefficients $\rho_j^{\alpha}(0)$: 
\begin{equation}
	\label{GeneralProbabilityCoeff}
	P_{\nu_\alpha\rightarrow\nu_\beta}=  
	\frac{1}{3}+\frac{1}{2}\sum_{i,j} \rho^{\beta}_{i}(0) \rho^{\alpha}_{j}(0) [e^{{\bf M}t}]_{ij} ,
	\end{equation}	
where $\beta,\alpha=e,\mu,\tau$ and $i,j=1,...,8$. 

\subsection{ The $\rho_j^{\alpha}(0)$ coefficients and Majorana phases}

The coefficients $\rho_j^{\alpha}(0)$ and $\rho_j^{\beta}(0)$ encloses the elements of the neutrino mixing matrix, as shown in Appendix \ref{AppendixA}. Thus, in order to include the Majorana phases, it is enough to make the replacement:  
\begin{equation}
 {U}_{\mathrm{Majorana}}=U_{\mathrm{PMNS}}.\mathrm{diag}(1,e^{-i \phi_1},e^{-i\phi_2})
\end{equation}
where $\phi_1$ and $\phi_2$ are the well-known Majorana phases. Therefore, the 
coefficients, when the Majorana phases are included, takes the following form: 
%corresponding ${\rho}_j^{\alpha}${'}s are described by the next equations: 
\begin{equation}
	\begin{split}		
       {\rho}_1^{\alpha}
&\rightarrow\rho_1^{\alpha}\cos{\phi_1}-\rho_2^{\alpha}\sin{\phi_1}\\
    {\rho}_2^{\alpha}
&\rightarrow\rho_2^{\alpha}\cos{\phi_1}+\rho_1^{\alpha}\sin{\phi_1}\\
    {\rho}_3^{\alpha}&\rightarrow\rho^{\alpha}_3\\
    {\rho}_4^{\alpha}
&\rightarrow\rho_4^{\alpha}\cos{\phi_2}
                        -\rho_5^{\alpha}\sin{\phi_2}\\
    {\rho}_5^{\alpha}
&\rightarrow\rho_5^{\alpha}\cos{\phi_2}
                        +\rho_4^{\alpha}\sin{\phi_2}\\
    {\rho}_6^{\alpha}    
&\rightarrow\rho_6^{\alpha}\cos{\Delta \phi}
                        -\rho_7^{\alpha}\sin{\Delta \phi}\\
    {\rho}_7^{\alpha}
&\rightarrow\rho_7^{\alpha}\cos{\Delta \phi}
                        +\rho_6^{\alpha}\sin{\Delta \phi}\\
    {\rho}_8^{\alpha} &\rightarrow\rho^{\alpha}_8,
	\end{split}
\end{equation}
where ${\rho}_j^{\alpha}$ are the coefficients considering only 
the $U_{\mathrm{PMNS}}$ mixing elements. For other Majorana neutrino mixing matrix parametrizations, the value of the Majorana phases in the above equations can be reinterpreted, see Appendix \ref{AppendixA}.		

\section{Transition probability - Perturbative approach}
Our selected texture of decoherence matrix $\bf{D}$ in the mass vacuum basis (MVB), can be seen as composed by two matrices: one is a diagonal one with its all element equals ${\bf{D}^{d}} =-\Gamma \times \bf{\mathbb{I}}$. The other one, $\bf{D}^{nd}$, is composed by its off-diagonal part, having as non-null only the given $-\Gamma_{ij} (=-\Gamma_{ji})$ elements (the diagonal is zero). As a consequence from the latter, the decoherence matrix in the mass matter basis (MMB) $\bf{D}_{m}$, obtained after rotating the aforementioned one, is defined by two different matrices, i.e. ${\bf{D}_{m}} = \bf{D}^{d}_{m} + \bf{D}^{nd}_{m} $.
 The ${\bf{D}^{d}_m} =-\Gamma \times \bf{\mathbb{I}}$ is purely diagonal,  the ${\bf{D}^{d}}$ is unaltered by the rotation, while  $\bf{D}^{nd}_{m}$ is the rotated matrix of the non-diagonal matrix 
$\bf{D}^{nd}$ in the MVB. 

Considering, that the transition probability when neutrinos travel through matter is:
\begin{equation}
P_{\nu_{\alpha}\rightarrow\nu_{\beta}}=\frac{1}{3}+\frac{1}{2}({{\varrho}_m^{\beta}(0)})^T {\varrho}_m^{\alpha}(t),
\end{equation}
where  ${\varrho}_m^{\alpha}(t)=e^{{\bf{M}} t} {\varrho}_m^{\alpha}(0) $, with ${\bf {M}}={\bf H}_{m}+{\bf{D}}_{m}$. Being ${\bf H}_{m}$ the Hamiltonian is written in the MMB. Due to  ${{\bf{D}}^{d}_m}$ commutes with any matrix  we have:
\begin{equation}
{\varrho}_m^{\alpha} (t)= e^{-\Gamma t} e^{({\bf H_m + D}_{m}^{nd})t} {\varrho}_m^{\alpha}(0)=e^{-\Gamma t}{\varrho'}_m^{\alpha} (t)
\end{equation} 
Given that ${\varrho_m^{\alpha} (0)=\varrho'}_m^{\alpha}(0)$ we can rewrite the probability in the following way: 
\begin{equation}
P_{\nu_{\alpha}\rightarrow\nu_{\beta}}=\frac{1}{3}+\frac{1}{2}e^{-\Gamma t}({\varrho'}_m^{\beta}(0))^T{\varrho'}_m^{\alpha} (t).
\label{probmatterdeco}
\end{equation}
The solution of 
${\varrho'}_m^{\alpha} (t)$, which is based on a power series solution expanded in $\theta_{13}$, $\alpha_{\Delta }={\Delta m_{12}^2}/{\Delta m_{13}^2}$ and a single
$\bar{\Gamma}_{ij}=\Gamma_{ij} t$, is shown in the Appendix \ref{AppendixB}. It follows 
the procedure the given in~\cite{Carpio:2017nui}.

After assessing the magnitude of the CP-odd terms in the transition, probability caused, individually, by each one of the off-diagonal elements
$\Gamma_{ij}$ (i. e. those who activate the Majorana phases) we conclude that $\Gamma_{28}=-\Gamma/\sqrt{3}$ gives us the most significant deviation from the standard oscillation formulae. We achieve the latter by fixing the off-diagonal elements at their maximum allowed values and taking the diagonal ones equal to $-\Gamma$ \cite{Carrasco:2018sca}.
In the Appendix \ref{Appendixc} is displayed the correspondence between each off-diagonal elements, $\Gamma_{ij}$, its ability to turn on CP-odd or CPT-odd terms and its connection to either $\phi_1$, $\phi_2$, or $\Delta \phi =\phi_1-\phi_2$.

Therefore, taking the off-diagonal part of the decoherence matrix $\bf{D}_{nd}$ formed only  with non-null $-\Gamma_{28}$ %a decoherence matrix composed by 
%a diagonal matrix $\Gamma \times \mathbb{1}$ plus an off-diagonal matrix with a  
%non-null $\Gamma_{28}$
,the following semi-analytical 
perturbative $\nu_\mu \rightarrow \nu_e$ transition probability formula 
for  $\text{SO}$ plus decoherence ($\text{DE}$)
is obtained:

		\begin{figure}
			\includegraphics[scale=0.58]{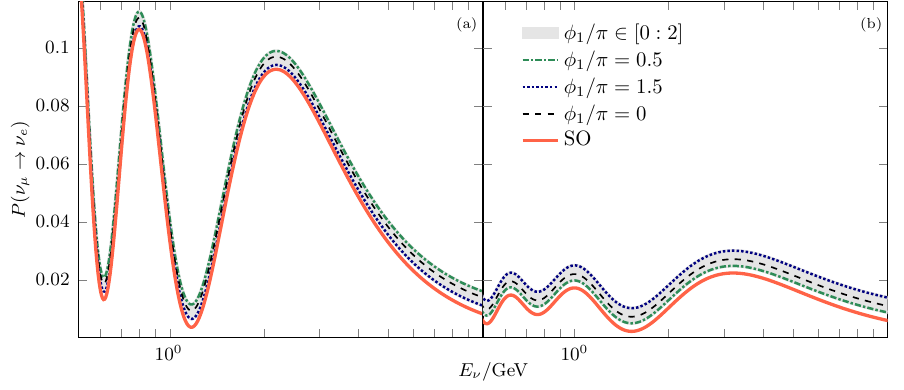}
			\caption{Oscillation probability depending on the neutrino energy for DUNE experiment. The figures ($\mathrm{a}$) and ($\mathrm{b}$) represent the $\nu_\mu \rightarrow \nu_e$ and $\bar{\nu}_\mu \rightarrow \bar{\nu}_e$ appearance channels respectively.  The off-diagonal decoherence parameter is  $\Gamma_{28}=-\Gamma/\sqrt{3}$. We consider $\Gamma_{28}=-\Gamma/\sqrt{3}$, $\delta_{\mathrm{CP}}/\pi=1.4$, and $\Gamma = 2.5\times 10^{-24} \mathrm{GeV}$.}
			\label{probabilities}
		\end{figure}

\begin{equation}
		\begin{split}
		&P^{\text{SO}\bigoplus \text{DE}}_{\nu_\mu \rightarrow \nu_e} = \frac{\big(1-e^{-\bar{\Gamma} }\big)}{3}+P_{\nu_\mu \rightarrow \nu_e}^{\text{SO}}e^{-\bar{\Gamma}}-\frac{\bar{\Gamma}_{28}}{\sqrt{3}}\sin{2{\theta_{12}}}\sin^2{\theta_{23}}\\ &\times \sin{\phi_1}e^{-\bar{\Gamma}}+\bar{\Gamma}_{28}\theta_{13}\frac{\sin{2\theta_{23}}}{2\sqrt{3}(A-1)A\Delta}\bigg(\big((1-A^2)\cos{\delta}\\
		&+A^2\cos{(\delta-\Delta)}-\cos{(\delta-A\Delta)}\big)\cos{\phi_1}+\big((1-A^2)\sin{\delta}\\
		&+A^2\sin{(\delta-\Delta)}-\sin{(\delta-A\Delta)}\big)\cos{2\theta_{12}}\sin{\phi_1}\bigg)e^{-\bar{\Gamma}}\\
		&+\bar{\Gamma}_{28}\alpha_\Delta\frac{\sin{2{\theta_{12}}}}{\sqrt{3}A^2\Delta}\bigg(2\sin^2{\bigg(\frac{A\Delta}{2}\bigg)}
		\cos{2\theta_{23}}\cos{\phi_1}-\cos{2\theta_{12}}\\
		&\times\sin^2{\theta_{23}}\big(\sin{A\Delta}-A\Delta\big)\sin{\phi_1}\bigg)e^{-\bar{\Gamma}}+... \\
		\end{split}
		\label{probnumunue}
		\end{equation}
where $P_{\nu_\mu \rightarrow \nu_e}^{\text{SO}}$ is the SO probability in matter ($t\rightarrow L$), $\Delta =(m^2_3 - m^2_1)
L/(2E)$ and $A=\sqrt{2}G_F n_e L/\Delta$, where $G_F$ is the Fermi constant, $n_e$ is the electron number density and $L$ is the source-detector distance. The validity of the transition probability formula relies on having the parameter-perturbative expansion
$\bar{\Gamma}_{28}$, $\bar{\Gamma}_{28}\theta_{13}$ 
and $\bar{\Gamma}_{28}\alpha_{\Delta}$ of 
order: $10^{-2}$, $10^{-3}$, and $10^{-4}$, respectively. For instance,  
the latter values can be attained for  
 $|\Gamma_{28}|=2\times 10^{-24}$ GeV, and at DUNE baseline $L=1300$ km. 
Furthermore, to get the antineutrino transition probability 
is enough to: $\phi_1 \rightarrow -\phi_1$, $\delta \rightarrow -\delta$ 
and $A \rightarrow -A$, meanwhile, the $\nu_\mu$ survival probability formula 
is not shown due to its negligible decoherence effect.

The $\nu_\mu \rightarrow \nu_e$ transition probability displayed in Fig.~\ref{probabilities} is numerically calculated at DUNE baseline and for the maximum value of $\Gamma_{28}=-\Gamma/\sqrt{3}$,  with $\Gamma=2.5 \times 10^{-24}$ GeV and the following values for the SO parameters, taken from~\cite{Nufit}: $\theta_{12}=33.82^{\circ}$, $\theta_{13} =8.61^{\circ}$, $\theta_{23} =48.3^{\circ}$, $\Delta m_{21}^{2} =7.39 \times 10^{-5} \mathrm{eV}^{2}$, and $\Delta m_{31}^{2} =2.523 \times 10^{-3} \mathrm{eV}^{2}$ (normal hierarchy), that are going to be fixed along this letter. 
The Dirac CP phase is taken as: $\delta_{\mathrm{CP}}^{true}/\pi=1.4$ 
inspired in the hint given by the T2K experiment
\cite{Abe:2019vii}
 From this figure it is notorious the energy-independent 
increase of the $\text{SO}\bigoplus \text{DE}$ probability respect to standard one, regardless the value of $\phi_1$, a feature that has been already pointed out in~\cite{Carpio:2017nui,Carpio:2018gum}, for other shape of the decoherence matrix. However, the intensity of this increment depends  on $\phi_1$, for example, in case of $\phi_1/ \pi=1.5$ ($\phi_1/ \pi=0.5$)  the $\text{SO}\bigoplus \text{DE}$ neutrino (antineutrino) probability grows much less than its antineutrino (neutrino) counterpart. For $\phi_1/ \pi=0$  the gain is proportionally the same for both, neutrinos or antineutrinos.

		\begin{figure}
			\includegraphics[scale=0.90]{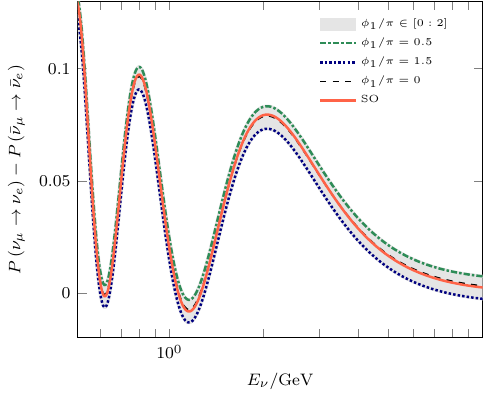}
			\caption{CP asymmetry depending on the neutrino energy. The off-diagonal decoherence parameter is  $\Gamma_{28}=-\Gamma/\sqrt{3}$. We consider $\delta_{\mathrm{CP}}/\pi=1.4$, and $\Gamma = 2.5\times 10^{-24} \mathrm{GeV}$.}
			\label{CPasymmetry}
		\end{figure}

In order to quantify the CP violating effects from the extra terms containing the Majorana phase given in our perturbatives formulae, we use the CP violation asymmetry $\Delta P=P_{\nu_\mu \rightarrow \nu_e} -P_{\bar{\nu}_\mu \rightarrow \bar{\nu}_e}$: 
\begin{equation}
\Delta P^{\text{SO}\bigoplus \text{DE}}\simeq
\Delta P^{\text{SO}} e^{-\bar{\Gamma}}
+\frac{2\bar{\Gamma}}{3}\sin{2{\theta_{12}}}\sin^2{\theta_{23}} \sin{\phi_1}e^{-\bar{\Gamma}}+...
\label{dp}
\end{equation}
here it is displayed only the leading term $\bar{\Gamma}_{28} \sim
\cal{O}$ (0.01) taken $\Gamma_{28}=-\Gamma/\sqrt{3}$, which is its maximum allowed value. The predictions
from
Eq. (\ref{dp}) are illustrated in Fig.~\ref{CPasymmetry}
where the $\nu_\mu \rightarrow \nu_e (\bar{\nu}_\mu \rightarrow \bar{\nu}_e)$ transition probability is numerically calculated at DUNE baseline for $\Gamma_{28}=-\Gamma/\sqrt{3}$,  with $\Gamma=2.5 \times 10^{-24}$ GeV. In  Fig.~\ref{CPasymmetry}  we see how the overall negative (positive) sign of the
decoherence contribution for
$\phi_1/ \pi=1.5$ ($\phi_1/ \pi=0.5$) diminish (increases) the
$\Delta P$ amplitude, whilst for $\phi_1/ \pi=0.0$ is, as expected, nearly equal to the SO case.

\section{Simulation and Results}
\begin{figure}
    \includegraphics[scale=0.55]{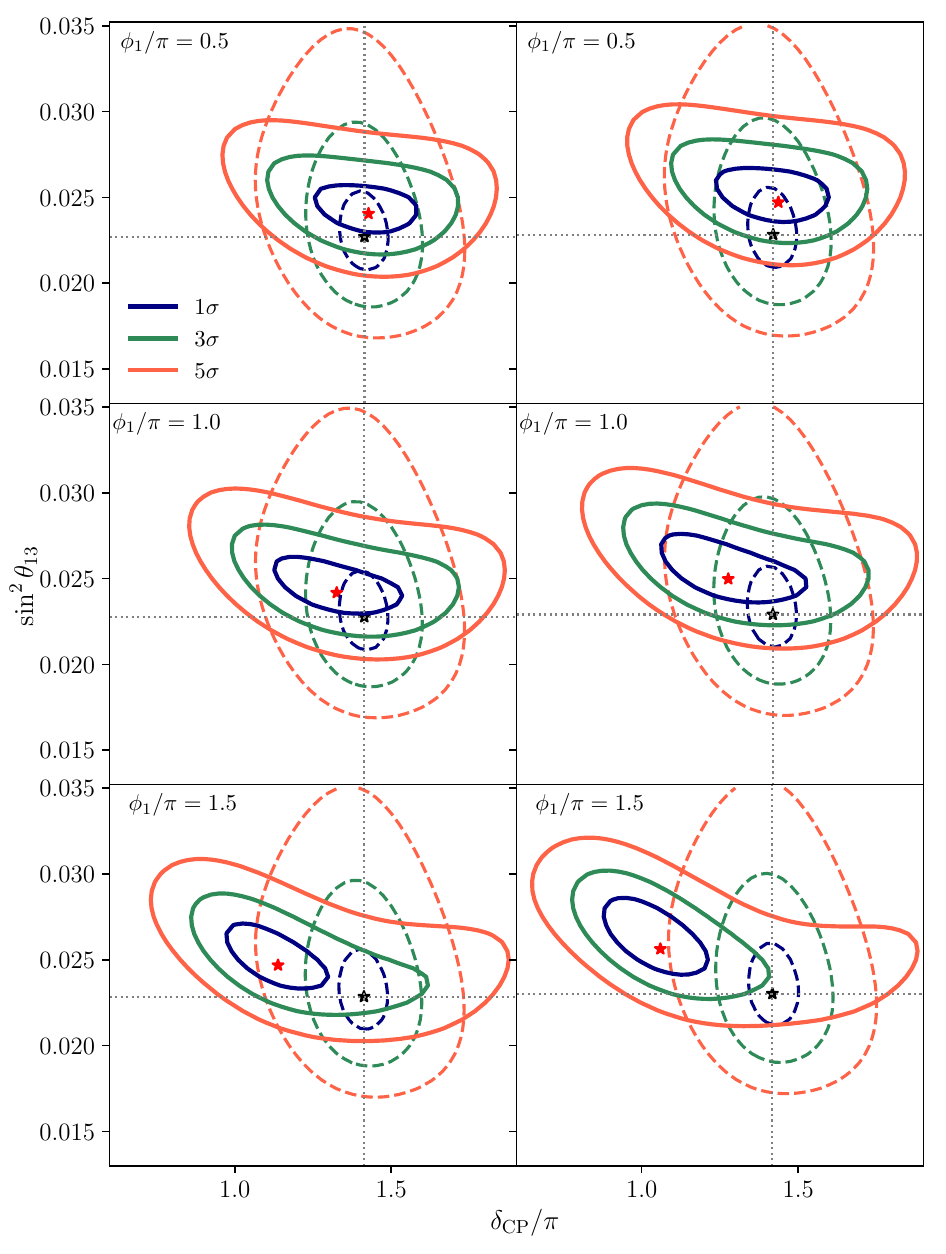}
    \caption{$\Delta \chi^2 $ contours (2 dof) considering the effects of decoherence with Majorana phases on the standard oscillation fits. The solid  and dashed lines are decoherence with $\Gamma_{28}=-\Gamma/\sqrt{3}$ for the DUNE and T2HK experiments, respectively. The left column is $\Gamma = 2.5\times 10^{-24} \mathrm{GeV}$ and the right column is $\Gamma = 3.5\times 10^{-24} \mathrm{GeV}$. We consider  $\delta_{\mathrm{CP}}^{true}/\pi=1.4$.}
    \label{dcp_vs_th13_panel}
\end{figure}

%\section{Experiment, Simulation and Results}
The DUNE and T2HK simulated data samples are generated 
with GLoBES~\cite{Huber:2004ka,Huber:2007ji} and nuSQuIDS~\cite{Delgado:2014kpa} 
introducing the configuration and inputs from \cite{Alion:2016uaj,Abe:2018uyc,Acciarri:2015uup,Huber:2007em}, and 
selecting the optimized fluxes for neutrino and antineutrino with $5$ years of exposure time per each mode  for DUNE with a 40-kt detector. While for T2HK, with 258-kt detector, we consider $3$ and $9$ years for neutrino and antineutrino mode, respectively. These simulated samples are created for non-null values of $\Gamma^{true}$ and $\phi_1^{true}$ 
and for a value of the Dirac CP violation phase set on the 
measurement performed by the T2K experiment: $\delta_{\mathrm{CP}}^{true}/\pi=1.4$ \cite{Abe:2019vii}. At this point it is important to mention that due to the small statistics and the large size of the uncertainties, we disregard the measurement of the Dirac CP violation claimed by NOvA experiment, which is $\delta_{\mathrm{CP}}/\pi\sim 0.82$ \cite{novaresults}.
In this analysis, the T2K measurement is considered as the true value of the Dirac CP violation phase since it should be unaltered by
any quantum decoherence effects. This is because of the small size of 
the higher decoherence contributions that would be 
$\bar{\Gamma} \sim \cal{O} \text{(0.001)}$, a consequence of combining  the source-detector distance of the T2K experiment with the $\Gamma$ 
elected for this study. It should be expected, that the T2HK experiment, with the same source-detector distance, would be also unaffected by the quantum decoherence effects. Within our analysis, the T2HK Dirac CP violation phase simulated measurement, which is an upgrade in the precision of the one performed at T2K, will be used as a reference point  with the expectations at DUNE. The 
$\chi^2$ analysis for DUNE and T2HK relies on the comparison between the SO phenomena, adopted as theoretical hypothesis, and simulated data that incorporates the quantum decoherence effects, where the 
prescription given in~\cite{Carpio:2018gum,Diaz:2020aax}
is followed. The calculation of the $\Delta \chi^2$ is described by:
\begin{equation}
\small
\begin{split}
   \Delta \chi^2=& \chi^2( \theta_{13}^{test}, \delta_{\mathrm{CP}}^{test}; \theta_{13}^{true}, \delta_{\mathrm{CP}}^{true}, \Gamma^{true},\phi_1^{true})\\ 
&- \chi_{min}^2( \theta_{13}^{fit}, \delta_{\mathrm{CP}}^{fit}; \theta_{13}^{true}, \delta_{\mathrm{CP}}^{true}, \Gamma^{true},\phi_1^{true})
\end{split}
\label{deltachi2}
\end{equation}
where $\theta_{13}^{fit}$ and $\delta_{\mathrm{CP}}^{fit}$ are 
the best-fit points which minimizes 
the $\chi^2$, considering priors at $3\sigma$ for the rest of the oscillation parameters but $\delta_{\mathrm{CP}}$. The DUNE and T2HK $\Delta \chi^2$ contours, projected into $\sin^2{\theta_{13}}$ vs $\delta_{\mathrm{CP}}$ planes and 
obtained after marginalizing over the rest of SO parameters, 
are presented in Fig. \ref{dcp_vs_th13_panel}. As 
expected, for T2HK, the $\sin^{2}{\theta^{fit}_{13}}$ and $\delta^{fit}_{\mathrm{CP}}$ are similar to the true ones being unmodified by the parameters chosen for decoherence. Meanwhile, for DUNE there is a slight increase of 
$\sin^{2}{\theta^{fit}_{13}}$, respect to the 
$\sin^{2}{\theta^{true}_{13}} (=0.0224)$, 
explicitly shown in Table.~\ref{tableresults}. This increment is the consequence of 
 trying to adjust the theoretical hypothesis, SO, with the 
energy-independent increase of the 
$\text{SO}\bigoplus \text{DE}$ probability amplitude embodied in the data, and modulated by the intensity of $\Gamma$ (see the third term of 
Eq.~\ref{probnumunue}). 

\begin{table}[!h]\begin{center}\begin{tabular}{|c|c|c|c|}
\hline
$\Gamma = 2.5 \times 10^{-24} \ \mathrm{GeV}$ & $\phi_1/\pi=0.5$ & $\phi_1/\pi=1.0$ & $\phi_1/\pi=1.5$ \\ \hline \hline
$\sin^2\theta_{13}^{fit}$ & $0.0241$ & $0.0242$ & $0.0247$ \\
$N_{\sigma}$ & $0.31\sigma$ & $0.34\sigma$ & $0.55\sigma$  \\		
\hline
$\delta_{\mathrm{CP}}^{fit}/\pi$ & $1.43$ & $1.33$ & $1.13$ \\
$N_{\sigma}$ & $0.08\sigma$ &  $1.19\sigma$ & $4.34\sigma$ \\
\hline	
$\Gamma = 3.5 \times 10^{-24} \ \mathrm{GeV}$ & $\phi_1/\pi=0.5$ & $\phi_1/\pi=1.0$ & $\phi_1/\pi=1.5$ \\  \hline 
$\sin^2\theta_{13}^{fit}$ & $0.0247$ & $0.0250$ & $0.0256$\\
$N_{\sigma}$ & $0.54\sigma$ & $0.61\sigma$ & $0.87\sigma$ \\		
\hline
$\delta_{\mathrm{CP}}^{fit}/\pi$ & $1.44$ & $1.28$ & $1.06$\\
$N_{\sigma}$ & $0.14\sigma$ &   $2.37 \sigma$ & $5.47\sigma$ \\	
\hline	
\end{tabular}
\caption{Fitted values for $\sin^2\theta_{13}$, $\delta_{\mathrm{CP}}$ and their respective shifts in terms of $\sigma$ units. We consider $\delta_{\mathrm{CP}}^{true}/\pi=1.4$}
\label{tableresults}
\end{center}\end{table}

The $\delta^{fit}_{\mathrm{CP}}$ for DUNE, when $ \phi_1/\pi= 1.5 $, 
is moving away 
from $\delta_{\mathrm{CP}}^{true}/\pi(=1.4)$ 
towards $\sim \pi$, minimizing the magnitude of 
the CP violation asymmetry. For $ \phi_1/\pi= 0.5$ the $\delta^{fit}_{\mathrm{CP}}$ takes almost exactly the
value of the true one going in the direction
to maximize 
the CP violation asymmetry.      
Both features, expressed numerically in Table.~\ref{tableresults}, can be explained from 
the need to accommodate the reduction (increase) of $\Delta P$,  
when $ \phi_1/\pi= 1.5 (0.5) $, seen in Fig.~\ref{CPasymmetry}. 
The quantified dislocation, in terms of $\sigma$, from $\sin^{2}{\theta^{fit}_{13}}$ and $\delta_{\mathrm{CP}}^{fit}$ (for DUNE) 
to the corresponding true ones (for T2HK), for 
 $\Gamma = \{2.5,3.5\}\times 10^{-24} \mathrm{GeV}$ and 
$\phi_1/\pi=\{0.0,0.5,1.5\} $, 
is depicted in Table~\ref{tableresults}. For $\Gamma =3.5(2.5)\times 10^{-24} \mathrm{GeV}$ 
the most prominent shift is found for 
$\phi_1/\pi=1.5$ with  $0.87 (0.55)\sigma$ and  $5.47 (4.34)\sigma$ for $\sin^{2}{\theta^{fit}_{13}}$ and $ \delta_{\mathrm{CP}}^{fit}$, respectively. While, for $\Gamma=3.5 \ (2.5) \times 10^{-24} \ \mathrm{GeV}$, the dislocation of $\delta_{\mathrm{CP}}^{fit}$ reaches  $3\sigma$ $(2\sigma)$ and $5\sigma$ $(3\sigma)$ when $\phi_1/\pi$ takes values below $1.01$ $(1.03)$ and $1.30$  $(1.10)$, respectively, the $\sin^{2}{\theta^{fit}_{13}}$ is clearly stable in front of changes along the $\phi_1$ interval.
The less significant distortion is for $\phi_1/\pi=0.5$ with  $0.54 (0.31)\sigma$ and $0.14 (0.08)\sigma$ for $\sin^{2}{\theta^{fit}_{13}}$ and $ \delta_{\mathrm{CP}}^{fit}$, respectively. 
A way to discriminate between different values of $\phi_1$ is 
through the ratio ($\cal R$) of the number of $\sigma$ deviation for $\sin^{2}{\theta^{fit}_{13}}$ to the corresponding ones for 
$\delta_{\mathrm{CP}}^{fit}$. In fact, a sort of discernment is achieved, for instance, for $\Gamma=3.5(2.5) \times 10^{-24} \ \mathrm{GeV}$, ${\cal R}\sim 0.16(0.13) - 0.26 (0.29)$ for 
the interval $\phi_1/\pi=1.0, 1.5$ reaching 
values up to $\sim 3.86(3.88)$ for $\phi_1/\pi=0.5$. A plus that 
reinforce the utility of $\cal R$ it is its low variations against changes of $\Gamma$. 

\begin{figure}
    \includegraphics[scale=0.58]{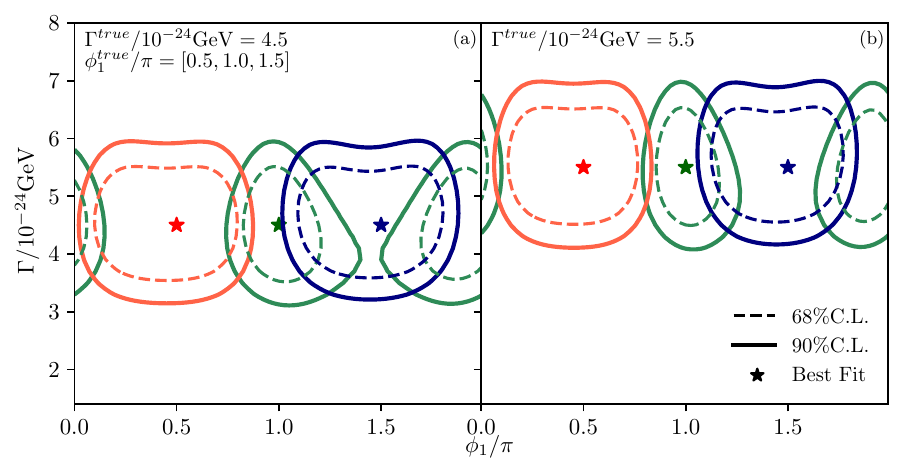}
    \caption{DUNE's ability to constrain (2 dof) the
decoherence parameter and the Majorana phase. The red, green and blue lines represent $\phi^{true}/\pi= 0.5,1.0,1.5$, respectively.}
    \label{panel_Majorana}
\end{figure}

The aforementioned analysis had the purpose of searching for distortions in 
$\sin^{2}{\theta^{fit}_{13}}$ and $\delta_{\mathrm{CP}}^{fit}$, considering pure SO as theoretical hypothesis. Now, the aim is to go one step
further and to explore the capacity of DUNE for 
measuring the Majorana phase, and also $\Gamma$, under the  
 ($\text{SO}$) plus decoherence ($\text{DE}$) as theoretical hypothesis, for $\phi_1/\pi= 0.5, 1, 1.5$ and for $\Gamma=4.5, 5.5 \times 10^{-24} \ \mathrm{GeV}$. The latter values cannot be compared with the limits imposed by Ice Cube \cite{Coloma:2018idr} since we are considering a non-diagonal scenario for the decoherence matrix.
In Fig.~\ref{panel_Majorana} the different allowed regions are displayed considering 68 $\%$ and 90 $\%$ C.L for 2 dof. Here it is seen that DUNE is able to measure $\phi_1/\pi = 1.50 \pm 0.35 (0.32)$ and  $\phi_1/\pi = 0.50 \pm 0.35 (0.32)$ 
and $\Gamma=4.50 \pm 1.38(5.50 \pm 1.42) \times 10^{-24} \ \mathrm{GeV}$. While for $\phi_1/\pi = 1.0 \pm 0.19 (0.15)$ a $\Gamma=4.50 \pm 1.42(5.50 \pm 1.46) \times 10^{-24} \ \mathrm{GeV}$ is obtained. It is interesting to 
remark that either for $\phi_1/\pi=0.5$ or $1.5$ the value of \(\pi\) is excluded at $90 \%$ C.L., independent the value of $\Gamma$. Thus, we are able to separate between these different values of $\phi_1$. 
%Putting in perspective with current results on the Dirac CPphase, it can be stated that the precision of $23 \%$ ($21 \%$) for the Majorana phase $\phi_1/\pi= 1.5$  with $\Gamma=4.5(5.5)\times 10^{-24} \mathrm{GeV}$, are rather competitive with the current one reached  in the determination of the Dirac CPphase of the T2K experiment \cite{Abe:2019vii}		
%, and are much better than the one achieved at NOVA \cite{novaresults}. Then, if decoherence exists in the manner is predicted here, there will be a very good chance that DUNE performs for the first time 
%the measurement of the Majorana CPphase 
%with not unreasonable uncertainties.  

\section{Summary and Conclusions}
We demonstrated the possibility of uncovering the Majorana nature of neutrinos in the DUNE experiment if a decoherence environment, with some determined characteristics, interacts with the neutrino system. Planck scale physics could be causing this decoherence environment.
Our approach is at first to show the strong displacement that it would be exhibited by the measured value of $\delta_{\mathrm{CP}}$ at DUNE, in comparison with the one measured at T2HK, which would be unaffected by  the decoherence effects. This displacement can be as large as
5.47 $\sigma$ for a Majorana phase $\phi_1/\pi= 1.5$ and a decoherence parameter
$\Gamma=3.5 \times 10^{-24} \ \mathrm{GeV}$. At next, we assessed the power of DUNE experiment in constraining the Majorana phase achieving, for instance, a precision of $23 \%$ ($21 \%$) for $\phi_1/\pi=1.5$
with $\Gamma=4.5(5.5)\times 10^{-24} \mathrm{GeV}$.
These values of precision are compatible with the current results on the Dirac CP phase reached by the T2K experiment \cite{Abe:2019vii}. Finally, we can conclude that, if decoherence exists in the manner is predicted here, there would be an interesting chance for DUNE to perform a first time a measurement of the Majorana CP phase, with some reasonable uncertainties.

\section{Acknowledgements}

 A. M. Gago acknowledges funding by the {\it Direcci\'on de Gesti\'on de la Investigaci\'on} at PUCP, through grants  DGI-2017-3-0019 and DGI 2019-3-0044. F. N. D\'iaz acknowledges CONCYTEC for the graduate  fellowship  under  Grant  No. 236-2015-FONDECYT. The authors also want to thank F. de Zela, J. Jones-P\'erez, J. L. Bazo, C. A. Arg\"uelles and  Mario A. Acero for useful suggestions and reading the manuscript.

\appendix
\section{Analysis of the $\rho^{\alpha}_{j}$ for different parametrizations}
\label{AppendixA}

The starting point of our analysis are the coefficients $\rho^{\alpha}_{j}$, where $j=0,1,...,8$ and $\alpha=e,\mu,\tau$, which are expressed in terms of the mixing matrix \cite{Gago:2002na}:
\begin{equation}
	\begin{split}
		\rho_0^{\alpha}&=\sqrt{2/3}\\
		\rho_1^{\alpha}&=2 \mathrm{Re} (U^*_{\alpha1}U_{\alpha2})\\
		\rho_2^{\alpha}&=-2 \mathrm{Im}(U^*_{\alpha1}U_{\alpha2})\\
		\rho_3^{\alpha}&=|U_{\alpha1}|^2-|U_{\alpha2}|^2\\
		\rho_4^{\alpha}&=2 \mathrm{Re}(U^*_{\alpha1}U_{\alpha3})\\
		\rho_5^{\alpha}&=-2 \mathrm{Im}(U^*_{\alpha1}U_{\alpha3})\\
		\rho_6^{\alpha}&=2 \mathrm{Re}(U^*_{\alpha2}U_{\alpha3})\\
		\rho_7^{\alpha}&=-2 \mathrm{Im}(U^*_{\alpha2}U_{\alpha3})\\
		\rho_3^{\alpha}&=\frac{1}{\sqrt{3}}(|U_{\alpha1}|^2+|U_{\alpha2}|^2-2|U_{\alpha3}|^2)\\
	\end{split}
	\label{Eq1}
\end{equation}
where the  $U_{\alpha j}$ are the matrix elements of the PMNS matrix ($U_{\mathrm{PMNS}}$) \cite{Maki:1962mu} without taking into account the Majorana phases. 

\subsection{Symmetrical Parametrization of the mixing matrix}
\label{AppendixA1}

The elements of symmetric parametrization of the mixing matrix, given in Eq. (5) in  \cite{Rodejohann:2011vc}, assuming the relation $\delta=\phi_{13}-\phi_{12}-\phi_{23}$, can  be written as follows: 

\begin{equation}
 \begin{split}
   U_{e 1} &\rightarrow U_{e 1} \\
   U_{e 2} &\rightarrow U_{e 2} e^{-i \phi_{12}}\\
   U_{e 3} &\rightarrow U_{e 3} e^{-i (\phi_{23}+ \phi_{12})}\\
   U_{\mu 1} &\rightarrow U_{\mu 1} e^{i \phi_{12}} \\
   U_{\mu 2} &\rightarrow U_{\mu 2}\\
   U_{\mu 3} &\rightarrow U_{\mu 3} e^{-i \phi_{23}}\\
   U_{\tau 1} &\rightarrow U_{\tau 1} e^{i (\phi_{23}+ \phi_{12})}\\
   U_{\tau 2} &\rightarrow U_{\tau 2} e^{i \phi_{23}}\\
   U_{\tau 3} &\rightarrow U_{\tau 3} \\
\end{split}
\end{equation} 

where $\phi_{13}$, $\phi_{12}$ and $\phi_{23}$ are the CP phases used in \cite{Rodejohann:2011vc}.

The corresponding ${\rho}_j^\alpha$  are described by the following relations: 
\begin{equation}
	\begin{split}		
       {\rho}_1^{\alpha}
&\rightarrow \rho_1^{\alpha}\cos{\phi_{12}}-\rho_2^{\alpha}\sin{\phi_{12}}\\
  {\rho}_2^{\alpha}
&=\rho_2^{\alpha}\cos{\phi_{12}}+\rho_1^{\alpha}\sin{\phi_{12}}\\
       {\rho}_3^{\alpha} &\rightarrow\rho^{\alpha}_3\\
     {\rho}_4^{\alpha}
&\rightarrow \rho_4^{\alpha}\cos{(\phi_{12}+\phi_{23})}
                        -\rho_5^{\alpha}\sin{(\phi_{12}+\phi_{23})}\\
{\rho}_5^{\alpha}
&\rightarrow \rho_5^{\alpha}\cos{(\phi_{12}+\phi_{23})}
                        +\rho_4^{\alpha}\sin{(\phi_{12}+\phi_{23})}\\
    {\rho}_6^{\alpha}
&\rightarrow \rho_6^{\alpha}\cos{\phi_{23}}
                        -\rho_7^{\alpha}\sin{\phi_{23}}\\
{\rho}_7^{\alpha}
&\rightarrow \rho_7^{\alpha}\cos{\phi_{23}}
                        +\rho_7^{\alpha}\sin{\phi_{23}}\\
   {\rho}_8^{\alpha} &\rightarrow \rho^{\alpha}_8,
	\end{split}
\end{equation}

where ${\rho}_j^\alpha$ are given in Eq. (\ref{Eq1})

\subsection{Particle data group parametrization type I: PDG I}
Here we analyze the mixing matrix parametrization given in \cite{Tanabashi:2018oca}, which includes the Majorana phases $\phi_1$ and $\phi_2$: 
\begin{equation}
	U_{\text{Majorana}}=U_{\text{PMNS}}.\text{diag}(\exp{i\phi_1},\exp{i\phi_2},1)
\end{equation} 
The corresponding ${\rho}_j^\alpha$ are described by the following equations: 
\begin{equation}
	\begin{split}		
       {\rho}_1^{\alpha}
&\rightarrow \rho_1^{\alpha}\cos{\Delta \phi}+\rho_2^{\alpha}\sin{\Delta \phi}\\
    {\rho}_2^{\alpha}
&\rightarrow \rho_2^{\alpha}\cos{\Delta \phi}-\rho_1^{\alpha}\sin{\Delta \phi}\\
    {\rho}_3^{\alpha}&\rightarrow \rho^{\alpha}_3\\
    {\rho}_4^{\alpha}
&\rightarrow \rho_4^{\alpha}\cos{\phi_1}
                        -\rho_5^{\alpha}\sin{\phi_1}\\
    {\rho}_5^{\alpha}
&\rightarrow \rho_5^{\alpha}\cos{\phi_1}
                        +\rho_4^{\alpha}\sin{\phi_1}\\
    {\rho}_6^{\alpha}    
&\rightarrow \rho_6^{\alpha}\cos{\phi_2}
                        -\rho_7^{\alpha}\sin{\phi_2}\\
    {\rho}_7^{\alpha}
&\rightarrow \rho_7^{\alpha}\cos{\phi_2}
                        +\rho_6^{\alpha}\sin{\phi_2}\\
    {\rho}_8^{\alpha} &\rightarrow \rho^{\alpha}_8,
	\end{split}
\end{equation}

Below, we present in table \ref{comparison} a summary of the equivalences between the different parameterizations.

\begin{table}[!h]\begin{center}\begin{tabular}{|c|c|c|}
\hline
Sym. $\leftrightarrow$ PDG I & Sym. $\leftrightarrow$ Our Work & PDG I $\leftrightarrow$ Our Work \\
\hline
$\phi_{12}+\phi_{23}\leftrightarrow\phi_1$ & $\ \ \ \ \ \ \ \ \phi_{12}\leftrightarrow\phi_1$ & $ \ \hspace{0.02cm} \phi_1 \leftrightarrow \Delta \phi$ \\
$\ \ \ \ \ \ \ \ \phi_{23}\leftrightarrow \phi_2$ & $\phi_{12}+\phi_{23} \leftrightarrow \phi_2$ & $ \phi_2 \leftrightarrow \phi_1$ \\
$\ \ \ \ \ \ \ \ \ \phi_{12}\leftrightarrow \Delta \phi$ & $\hspace{-0.05cm} \ \ \ \ \ \ \ \ \ \ \ \ \phi_{23}\leftrightarrow -\Delta\phi$ & $\ 
\hspace{0.02cm} \Delta \phi \leftrightarrow -\phi_{2}$\\	
\hline
\end{tabular}
\caption{Parameterizations comparison.}
\label{comparison}
\end{center}\end{table}

where $\Delta \phi = \phi_1 - \phi_2$.

%\subsubsection{Symmetric and PDG type I}
%\begin{equation}
%\begin{split}
%\phi_{12} \leftrightarrow & \Delta \phi \\
%(\phi_{12}+\phi_{23}) \leftrightarrow & \phi_1\\
%\phi_{23} \leftrightarrow & \phi_2
%\end{split}
%\end{equation}
%\subsubsection{Symmetric and PDG type II}
%\begin{equation}
%\begin{split}
%\phi_{12} \leftrightarrow & \phi_1 \\
%(\phi_{12}+\phi_{23}) \leftrightarrow & \phi_2 \\
%\phi_{23} \leftrightarrow & -\Delta \phi
%\end{split}
%\end{equation}
%\subsubsection{PDG type I and PDG type II}
%\begin{equation}
%\begin{split}
%\phi_1 \leftrightarrow & \Delta\phi \\
%\phi_2 \leftrightarrow & \phi_1 \\
%\Delta \phi \leftrightarrow &- \phi_2
%\end{split}
%\end{equation}

\section{Probability Calculation}
\label{AppendixB}

For solving ${\varrho'}_m^{\alpha} (t)$ we must start with the next differential equation: 
\begin{equation}
	\label{Evolutionequation}
	\dot{{\varrho'}_m^\alpha}= ({\bf H_m + D}_{m}^{nd}) 
{\varrho'}_m^\alpha,
	\end{equation}
which is similar to Eq.(2) presented in our letter. Before continue, we must point out that the following procedure is similar to the one given in \cite{Carpio:2017nui}.  The Eq.(\ref{Evolutionequation}) can be simplified using this change of variable: 
 \begin{equation}
{\varrho}_m^\alpha (t)=e^{{\bf H_{m}}t}{\tilde{\varrho}}^\alpha (t),
\label{eqq}
\end{equation} 
then the Eq. \ref{Evolutionequation} : 
\begin{equation}
e^{{\bf H_m}t}\dot{\tilde{\varrho}}^\alpha+
{\bf H}_m e^{{\bf H_m} t}
\tilde{\varrho}^\alpha={\bf (H_m+D_m^{nd})} e^{{\bf H_m} t}{\tilde{\varrho}}^\alpha 
\end{equation}
thus we get:
\begin{equation}
\dot{\tilde{\varrho}}^\alpha =e^{- {\bf H_m} t} {\bf D_m^{nd}} e^{- {\bf H_m} t}\tilde{\varrho}^\alpha,
\label{identity}
\end{equation}
the matrix $e^{- {\bf H_m} t} {\bf D_m^{nd}} e^{- {\bf H_m} t}$ 
can be expanded perturbatively in power series of the small parameters 
$\theta_{13}$, and $\alpha_\Delta$ which turns out to be: 
\begin{equation}
e^{- {\bf H_m} t} {\bf D_m^{nd}} e^{- {\bf H_m} t}
=\Gamma_{ij} (\tilde{D}^{(0)}+\theta_{13}\tilde{D}^{(\theta_{13})}+\alpha_\Delta \tilde{D}^{(\alpha_\Delta)}+...)
\label{identity2}
\end{equation}
we can factor out the decoherence parameter $\Gamma_{ij}$ 
since it is a common factor of all the elements in the decoherence matrix
$\bf D_m^{nd}$ in the MMB (a consequence of its definition in the 
mass vacuum basis that is an off-diagonal matrix with only non-null terms
in a given $-\Gamma_{ij}$ element). Replacing Eq.(\ref{identity2}) into Eq.(\ref{identity}):
\begin{equation}
\dot{\tilde{\varrho}}^\alpha=\Gamma_{ij} (\tilde{D}^{(0)}+\theta_{13}\tilde{D}^{(\theta_{13})}+\alpha_\Delta \tilde{D}^{(\alpha_\Delta)}+...) \tilde{\varrho}^\alpha
\label{diff1}
\end{equation}
the above equation can be solved perturbatively treating $\tilde{\varrho}^\alpha$ as a power series in $\theta_{13}$, $\alpha_\Delta$ and $\Gamma_{ij}$:
\begin{equation}
\begin{split}
\tilde{\varrho}^\alpha&=\tilde{\varrho}^{(0)}+\theta_{13}\tilde{\varrho}^{(\theta)}+
\alpha_\Delta \tilde{\varrho}^{(\alpha_\Delta)}+\alpha_\Delta \theta_{13}\tilde{\varrho}^{(\alpha_\Delta \theta_{13})}+...\\
&+\Gamma_{ij} \tilde{\varrho}^{(\Gamma_{ij})}
 +\Gamma_{ij} \theta_{13} \tilde{\varrho}^{(\Gamma_{ij}\theta_{13})}
+\Gamma_{ij} \alpha_\Delta  \tilde{\varrho}^{(\Gamma_{ij}\alpha_\Delta)}
+...
\end{split}
\label{rho1}
\end{equation}
Then substituing Eq.(\ref{rho1}) into Eq.(\ref{diff1}) we produce a sequence of first order differential equations each of them collecting 
equal power terms. The $\Gamma_{ij}$-independent terms of the
$\tilde{\varrho}^\alpha$ expansion: $\tilde{\varrho}^{(0)}+\theta_{13}\tilde{\varrho}^{(\theta)}+
\alpha_\Delta \tilde{\varrho}^{(\alpha_\Delta)}+\alpha_\Delta \theta_{13}\tilde{\varrho}^{(\alpha_\Delta \theta_{13})}+...\ $    
corresponds to the initial condition $\tilde{\varrho}^\alpha (0)$, which is constant in time and coincides with the initial condition for the standard oscillation case, since at that instant the environment is decoupled (not interacting) with the neutrino system. Considering all the later plus the condition that ${\varrho'}^\alpha_m(0)=\tilde{\varrho}^\alpha(0)$, we can rewrite Eq.(\ref{eqq}) as follows:
\begin{equation}
{\varrho'}_m^\alpha (t)=e^{{\bf H_{m}}t}({\varrho'}_m^\alpha (0)
+\bar{\Gamma}_{ij}(...)). ,
\label{eqq2}
\end{equation} 
with $\bar{\Gamma}_{ij}=\Gamma_{ij} t$. The second term at the right-hand side of the equation above contains the explicit solution 
of the power series of ${\varrho}^\alpha$.

\section{CP-odd, CPT-odd terms and Majorana phases}
\label{Appendixc}

\begin{table}[!h]\begin{center}\begin{tabular}{|c|c|c|}
\hline
$\mathrm{CPV}$ & $\mathrm{CPTV}$ & Non-null Majorana phase  \\\hline \hline
$\Gamma_{13},\Gamma_{23},\Gamma_{18},\Gamma_{28},\Gamma_{12}$& $\Gamma_{23},\Gamma_{28},\Gamma_{12}$ &$\phi_{1}$\\

$\Gamma_{34},\Gamma_{35},\Gamma_{48},\Gamma_{58},\Gamma_{45}$&$\Gamma_{35},\Gamma_{58},\Gamma_{45}$&$\phi_{2}$  \\

$\Gamma_{37},\Gamma_{36},\Gamma_{68},\Gamma_{67},\Gamma_{78}$ & $\Gamma_{37},\Gamma_{67},\Gamma_{78}$ & $\Delta\phi$ \\

$\Gamma_{14},\Gamma_{24},\Gamma_{15},\Gamma_{25}$ &$\Gamma_{24},\Gamma_{15}$ &$\phi_{1}, \phi_{2}$ \\

$\Gamma_{16},\Gamma_{17},\Gamma_{26},\Gamma_{27}$ &$\Gamma_{17},\Gamma_{26}$ & $\phi_{1}, \Delta \phi$ \\

$\Gamma_{46},\Gamma_{47},\Gamma_{56},\Gamma_{57}$ & $\Gamma_{47},\Gamma_{56}$ & $\phi_{2}, \Delta \phi$ \\		
\hline	
\end{tabular}
\caption{Violation of symmetries by non-diagonal decoherence elements and their dependence on the Majorana phases.}
\label{sas}
\end{center}\end{table}
In table~\ref{sas} we present a classification of the correspondence between each one of the off-diagonal elements $\Gamma_{ij}$  and $\phi_{1}$, $\phi_{2}$, or $\Delta \phi$,  also pointing out its connection with CP-odd, CPT-odd terms or both in the oscillation probabilities which incorporates quantum decoherence.


\begin{thebibliography}{0}
\expandafter\ifx\csname natexlab\endcsname\relax\def\natexlab#1{#1}\fi
\expandafter\ifx\csname bibnamefont\endcsname\relax
  \def\bibnamefont#1{#1}\fi
\expandafter\ifx\csname bibfnamefont\endcsname\relax
  \def\bibfnamefont#1{#1}\fi
\expandafter\ifx\csname citenamefont\endcsname\relax
  \def\citenamefont#1{#1}\fi
\expandafter\ifx\csname url\endcsname\relax
  \def\url#1{\texttt{#1}}\fi
\expandafter\ifx\csname urlprefix\endcsname\relax\def\urlprefix{URL }\fi
\providecommand{\bibinfo}[2]{#2}
\providecommand{\eprint}[2][]{\url{#2}}

\end{thebibliography}


\begin{thebibliography}{}


%\cite{Murayama:2006qb}
\bibitem{Murayama:2006qb}
H.~Murayama,
%``Origin of neutrino mass,''
\href{https://www.sciencedirect.com/science/article/abs/pii/S0146641006000214?via%3Dihub}{Prog. Part. Nucl. Phys. \textbf{57}, 3-21 (2006)}
%doi:10.1016/j.ppnp.2006.02.001
%2 citations counted in INSPIRE as of 22 Feb 2021



%\cite{Valle:2015mma}
\bibitem{Valle:2015mma}
J.~W.~F.~Valle,
%``Status and implications of neutrino masses: a brief panorama,''
\href{https://www.worldscientific.com/doi/abs/10.1142/S0217751X15300343}{Adv. Ser. Direct. High Energy Phys. \textbf{25}, 25-37 (2015)}
%doi:10.1142/S0217751X15300343
%[arXiv:1504.01913 [hep-ph]].
%4 citations counted in INSPIRE as of 22 Feb 2021


%\cite{Yanagida:1979as}
\bibitem{Yanagida:1979as}
T.~Yanagida,
%``Horizontal gauge symmetry and masses of neutrinos,''
\href{https://inspirehep.net/literature/143150}{Conf. Proc. C \textbf{7902131}, 95-99 (1979)}
%KEK-79-18-95.
%1903 citations counted in INSPIRE as of 22 Feb 2021


%\cite{GellMann:1980vs}
\bibitem{GellMann:1980vs}
M.~Gell-Mann, P.~Ramond and R.~Slansky,
%``Complex Spinors and Unified Theories,''
\href{https://arxiv.org/abs/1306.4669}{Conf. Proc. C \textbf{790927}, 315-321 (1979)}
%[arXiv:1306.4669 [hep-th]].
%3301 citations counted in INSPIRE as of 22 Feb 2021

%\cite{Glashow:1979nm}
\bibitem{Glashow:1979nm} 
  S.~L.~Glashow,
  %``The Future of Elementary Particle Physics,''
  \href{https://link.springer.com/chapter/10.1007%2F978-1-4684-7197-7_15}{NATO Sci.\ Ser.\ B {\bf 61}, 687 (1980).}
  %doi:10.1007/978-1-4684-7197-7_15
  %%CITATION = doi:10.1007/978-1-4684-7197-7_15;%%
  %484 citations counted in INSPIRE as of 22 Feb 2021

%\cite{Minkowski:1977sc}
\bibitem{Minkowski:1977sc}
P.~Minkowski,
%``$\mu \to e\gamma$ at a Rate of One Out of $10^{9}$ Muon Decays?,''
\href{https://doi.org/10.1016/0370-2693(77)90435-X}{Phys. Lett. B \textbf{67}, 421-428 (1977)}
%doi:10.1016/0370-2693(77)90435-X
%4020 citations counted in INSPIRE as of 19 Feb 2021


%
\bibitem{Mohapatra:1979ia}
R.~N.~Mohapatra and G.~Senjanovic,
%``Neutrino Mass and Spontaneous Parity Nonconservation,''
\href{https://journals.aps.org/prl/abstract/10.1103/PhysRevLett.44.912}{Phys. Rev. Lett. \textbf{44}, 912 (1980)}
%doi:10.1103/PhysRevLett.44.912
%5595 citations counted in INSPIRE as of 18 Feb 2021

%\cite{Magg:1980ut}
\bibitem{Magg:1980ut}
M.~Magg and C.~Wetterich,
%``Neutrino Mass Problem and Gauge Hierarchy,''
\href{https://doi.org/10.1016/0370-2693(80)90825-4}{Phys. Lett. B \textbf{94}, 61-64 (1980)}
%doi:10.1016/0370-2693(80)90825-4
%975 citations counted in INSPIRE as of 18 Feb 2021



%\cite{Schechter:1980gr}
\bibitem{Schechter:1980gr}
J.~Schechter and J.~W.~F.~Valle,
%``Neutrino Masses in SU(2) x U(1) Theories,''
\href{https://journals.aps.org/prd/abstract/10.1103/PhysRevD.22.2227}{Phys. Rev. D \textbf{22}, 2227 (1980)}
%doi:10.1103/PhysRevD.22.2227
%2940 citations counted in INSPIRE as of 18 Feb 2021

%\cite{Mohapatra:1980yp}
\bibitem{Mohapatra:1980yp}
R.~N.~Mohapatra and G.~Senjanovic,
%``Neutrino Masses and Mixings in Gauge Models with Spontaneous Parity Violation,''
\href{https://journals.aps.org/prd/abstract/10.1103/PhysRevD.23.165}{Phys. Rev. D \textbf{23}, 165 (1981)}
%doi:10.1103/PhysRevD.23.165
%2627 citations counted in INSPIRE as of 19 Feb 2021


%%%%%%%%%%%%%%%%%%%%%%%%%%%%%%%%%%Seesaw%%%%%%%%%%%%%%%%%%%%%%%%%%%

%\cite{Foot:1988aq}
\bibitem{Foot:1988aq}
R.~Foot, H.~Lew, X.~G.~He and G.~C.~Joshi,
%``Seesaw Neutrino Masses Induced by a Triplet of Leptons,''
\href{https://link.springer.com/article/10.1007%2FBF01415558}{Z. Phys. C \textbf{44}, 441 (1989)}
%doi:10.1007/BF01415558
%903 citations counted in INSPIRE as of 19 Feb 2021



%\cite{Shrock:1980ct}
\bibitem{Shrock:1980ct}
R.~E.~Shrock,
%``General Theory of Weak Leptonic and Semileptonic Decays. 1. Leptonic Pseudoscalar Meson Decays, with Associated Tests For, and Bounds on, Neutrino Masses and Lepton Mixing,''
\href{https://journals.aps.org/prd/abstract/10.1103/PhysRevD.24.1232}{Phys. Rev. D \textbf{24}, 1232 (1981)}
%doi:10.1103/PhysRevD.24.1232
%357 citations counted in INSPIRE as of 19 Feb 2021


%\cite{Konetschny:1977bn}
\bibitem{Konetschny:1977bn}
W.~Konetschny and W.~Kummer,
%``Nonconservation of Total Lepton Number with Scalar Bosons,''
\href{https://doi.org/10.1016/0370-2693(77)90407-5}{Phys. Lett. B \textbf{70}, 433-435 (1977)}
%doi:10.1016/0370-2693(77)90407-5
%355 citations counted in INSPIRE as of 19 Feb 2021



%\cite{Cheng:1980qt}
\bibitem{Cheng:1980qt}
T.~P.~Cheng and L.~F.~Li,
%``Neutrino Masses, Mixings and Oscillations in SU(2) x U(1) Models of Electroweak Interactions,''
\href{https://journals.aps.org/prd/abstract/10.1103/PhysRevD.22.2860}{Phys. Rev. D \textbf{22}, 2860 (1980)}
%doi:10.1103/PhysRevD.22.2860
%1050 citations counted in INSPIRE as of 19 Feb 2021



%\cite{Lazarides:1980nt}
\bibitem{Lazarides:1980nt}
G.~Lazarides, Q.~Shafi and C.~Wetterich,
%``Proton Lifetime and Fermion Masses in an SO(10) Model,''
\href{https://doi.org/10.1016/0550-3213(81)90354-0}{Nucl. Phys. B \textbf{181}, 287-300 (1981)}
%doi:10.1016/0550-3213(81)90354-0
%1433 citations counted in INSPIRE as of 19 Feb 2021


%%%%%%%%%%%%%%%%%%%%%%Leptop Number Violation%%%%%%%%%%%%%%%%%%%%%%


%\cite{Atre:2009rg}
\bibitem{Atre:2009rg}
A.~Atre, T.~Han, S.~Pascoli and B.~Zhang,
%``The Search for Heavy Majorana Neutrinos,''
\href{https://iopscience.iop.org/article/10.1088/1126-6708/2009/05/030}{JHEP \textbf{05}, 030 (2009)}
%doi:10.1088/1126-6708/2009/05/030
%[arXiv:0901.3589 [hep-ph]].
%631 citations counted in INSPIRE as of 19 Feb 2021



%\cite{Zee:1985id}
\bibitem{Zee:1985id}
A.~Zee,
%``Quantum Numbers of Majorana Neutrino Masses,''
\href{https://www.sciencedirect.com/science/article/abs/pii/055032138690475X?via%3Dihub}{Nucl. Phys. B \textbf{264}, 99-110 (1986)}
%doi:10.1016/0550-3213(86)90475-X
%457 citations counted in INSPIRE as of 19 Feb 2021


%\cite{Datta:1993nm}
%\bibitem{Datta:1993nm}
%A.~Datta, M.~Guchait and A.~Pilaftsis,
%``Probing lepton number violation via majorana neutrinos at hadron supercolliders,''
%\href{https://journals.aps.org/prd/abstract/10.1103/%PhysRevD.50.3195}{Phys. Rev. D \textbf{50}, 3195-3203 %(1994)}
%doi:10.1103/PhysRevD.50.3195
%[arXiv:hep-ph/9311257 [hep-ph]].
%171 citations counted in INSPIRE as of 19 Feb 2021









%%%%%%%%%%%%%%%%%%%%%%%%%Doble beta decay%%%%%%%%%%%%%%%%%%%%%%%%%%%%%%%


%\cite{Bilenky:2012qi}
\bibitem{Bilenky:2012qi}
S.~M.~Bilenky and C.~Giunti,
%``Neutrinoless double-beta decay: A brief review,''
\href{https://doi.org/10.1142/S0217732312300157}{Mod. Phys. Lett. A \textbf{27}, 1230015 (2012)}
%doi:10.1142/S0217732312300157
%[arXiv:1203.5250 [hep-ph]].
%170 citations counted in INSPIRE as of 19 Feb 2021




%
\bibitem{Faessler:1999zg}
A.~Faessler and F.~Simkovic,
%``Double beta decay,''
\href{https://iopscience.iop.org/article/10.1088/0954-3899/24/12/001}{J. Phys. G \textbf{24}, 2139-2178 (1998)}
%doi:10.1088/0954-3899/24/12/001
%[arXiv:hep-ph/9901215 [hep-ph]].
%284 citations counted in INSPIRE as of 19 Feb 2021

%\cite{Avignone:2007fu}
\bibitem{Avignone:2007fu}
Avignone~F.T., Elliott~S.R. and Engel~J.,
%``Double Beta Decay, Majorana Neutrinos, and Neutrino Mass,''
\href{https://journals.aps.org/rmp/abstract/10.1103/RevModPhys.80.481}{Rev. Mod. Phys. \textbf{80}, 481-516 (2008)}
%doi:10.1103/RevModPhys.80.481
%[arXiv:0708.1033 [nucl-ex]].
%860 citations counted in INSPIRE as of 19 Feb 2021


%\cite{Menendez:2008jp}
%\bibitem{Menendez:2008jp}
%J.~Menendez, A.~Poves, E.~Caurier and F.~Nowacki,
%``Disassembling the Nuclear Matrix Elements of the Neutrinoless beta beta Decay,''
%\href{https://doi.org/10.1016/j.nuclphysa.2008.12.005}{Nucl. Phys. A \textbf{818}, 139-151 (2009)}
%doi:10.1016/j.nuclphysa.2008.12.005
%[arXiv:0801.3760 [nucl-th]].
%460 citations counted in INSPIRE as of 19 Feb 2021


%\cite{Caurier:2007wq}
%\bibitem{Caurier:2007wq}
%E.~Caurier, J.~Menendez, F.~Nowacki and A.~Poves,
%``The Influence of pairing on the nuclear matrix elements of the neutrinoless beta beta decays,''
%\href{https://journals.aps.org/prl/abstract/10.1103/PhysRevLett.100.052503}{Phys. Rev. Lett. \textbf{100}, 052503 (2008)}
%doi:10.1103/PhysRevLett.100.052503
%[arXiv:0709.2137 [nucl-th]].
%271 citations counted in INSPIRE as of 19 Feb 2021



%\cite{Vergados:2012xy}
\bibitem{Vergados:2012xy}
J.~D.~Vergados, H.~Ejiri and F.~Simkovic,
%``Theory of Neutrinoless Double Beta Decay,''
\href{https://iopscience.iop.org/article/10.1088/0034-4885/75/10/106301}{Rept. Prog. Phys. \textbf{75}, 106301 (2012)}
%doi:10.1088/0034-4885/75/10/106301
%[arXiv:1205.0649 [hep-ph]].
%411 citations counted in INSPIRE as of 19 Feb 2021





%%%%%%%%%%%%%%%%%%%%%%%%%%%No Signal%%%%%%%%%%%%%%%%%%%%%%%%%%%%%%%%%%


%\cite{Agostini:2019hzm}
\bibitem{Agostini:2019hzm}
M.~Agostini \textit{et al.} [GERDA],
%``Probing Majorana neutrinos with double-$\beta$ decay,''
\href{https://science.sciencemag.org/content/365/6460/1445}{Science \textbf{365}, 1445 (2019)}
%doi:10.1126/science.aav8613
%[arXiv:1909.02726 [hep-ex]].
%70 citations counted in INSPIRE as of 22 Feb 2021


%\cite{Anton:2019wmi}
\bibitem{Anton:2019wmi}
G.~Anton \textit{et al.} [EXO-200],
%``Search for Neutrinoless Double-$\beta$ Decay with the Complete EXO-200 Dataset,''
\href{https://journals.aps.org/prl/abstract/10.1103/PhysRevLett.123.161802}{Phys. Rev. Lett. \textbf{123}, no.16, 161802 (2019)}
%doi:10.1103/PhysRevLett.123.161802
%[arXiv:1906.02723 [hep-ex]].
%58 citations counted in INSPIRE as of 22 Feb 2021




%\cite{Adams:2019jhp}
\bibitem{Adams:2019jhp}
D.~Q.~Adams \textit{et al.} [CUORE],
%``Improved Limit on Neutrinoless Double-Beta Decay in  $^{130}$Te with CUORE,''
\href{https://journals.aps.org/prl/abstract/10.1103/PhysRevLett.124.122501}{Phys. Rev. Lett. \textbf{124}, no.12, 122501 (2020)}
%doi:10.1103/PhysRevLett.124.122501
%[arXiv:1912.10966 [nucl-ex]].
%49 citations counted in INSPIRE as of 19 Feb 2021




%\cite{KamLAND-Zen:2016pfg}
\bibitem{KamLAND-Zen:2016pfg}
Gando~A., Gando~Y., Hachiya~T., Hayashi~A., Hayashida~S., Ikeda~H. \textit{et al.} [KamLAND-Zen],
%``Search for Majorana Neutrinos near the Inverted Mass Hierarchy Region with KamLAND-Zen,''
\href{https://journals.aps.org/prl/abstract/10.1103/PhysRevLett.117.082503}{Phys. Rev. Lett. \textbf{117}, no.8, 082503 (2016)}
%doi:10.1103/PhysRevLett.117.082503
%[arXiv:1605.02889 [hep-ex]].
%760 citations counted in INSPIRE as of 19 Feb 2021






%%%%%%%%%%%%%%%%%%%%%%%%%%%%%%%%Another Way %%%%%%%%%%%%%%%%%%%%%%%%%%%%%



%
\bibitem{Sirunyan:2018xiv}
A.~M.~Sirunyan \textit{et al.} [CMS],
%``Search for heavy Majorana neutrinos in same-sign dilepton channels in proton-proton collisions at $ \sqrt{s}=13 $ TeV,''
\href{https://link.springer.com/article/10.1007%2FJHEP01%282019%29122}{JHEP \textbf{01}, 122 (2019)}
%doi:10.1007/JHEP01(2019)122
%[arXiv:1806.10905 [hep-ex]].
%45 citations counted in INSPIRE as of 19 Feb 2021



%\cite{Aaij:2014aba}
\bibitem{Aaij:2014aba}
R.~Aaij \textit{et al.} [LHCb],
%``Search for Majorana neutrinos in $B^- \to \pi^+\mu^-\mu^-$ decays,''
\href{https://journals.aps.org/prl/abstract/10.1103/PhysRevLett.112.131802}{Phys. Rev. Lett. \textbf{112}, no.13, 131802 (2014)}
%doi:10.1103/PhysRevLett.112.131802
%[arXiv:1401.5361 [hep-ex]].
%132 citations counted in INSPIRE as of 19 Feb 2021



%\cite{Balantekin:2018ukw}
\bibitem{Balantekin:2018ukw}
A.~B.~Balantekin, A.~de Gouv\^ea and B.~Kayser,
%``Addressing the Majorana vs. Dirac Question with Neutrino Decays,''
\href{https://www.sciencedirect.com/science/article/pii/S0370269318310001?via%3Dihub}{Phys. Lett. B \textbf{789}, 488-495 (2019)}
%doi:10.1016/j.physletb.2018.11.068
%[arXiv:1808.10518 [hep-ph]].
%25 citations counted in INSPIRE as of 22 Feb 2021




%\cite{Bora:2016ygl}
\bibitem{Bora:2016ygl}
K.~Bora, D.~Borah and D.~Dutta,
%``Probing Majorana Neutrino Textures at DUNE,''
\href{https://journals.aps.org/prd/abstract/10.1103/PhysRevD.96.075006}{Phys. Rev. D \textbf{96}, no.7, 075006 (2017)}
%doi:10.1103/PhysRevD.96.075006
%[arXiv:1611.01097 [hep-ph]].
%11 citations counted in INSPIRE as of 22 Feb 2021



%%%%%%%%%%%%%%%%%%%%%%%%%%Phase Absorbed %%%%%%%%%%%%%%%%%%%%%%%%

%\cite{Giunti:2010ec}
\bibitem{Giunti:2010ec}
C.~Giunti,
%``No Effect of Majorana Phases in Neutrino Oscillations,''
\href{https://doi.org/10.1016/j.physletb.2010.02.020}{Phys. Lett. B \textbf{686}, 41-43 (2010)}
%doi:10.1016/j.physletb.2010.02.020
%[arXiv:1001.0760 [hep-ph]].
%27 citations counted in INSPIRE as of 19 Feb 2021


%%%%%%%%%%%%%%%%%%%%%%%%%%%%%%%%%%vDecoheherence %%%%%%%%%%%55		


		
\bibitem{Benatti:2000ph} 
  F.~Benatti and R.~Floreanini,
  %``Open system approach to neutrino oscillations,''
  \href{https://doi.org/10.1088/1126-6708/2000/02/032}{JHEP {\bf 0002}, 032 (2000)}; \href{https://doi.org/10.1103/PhysRevD.64.085015}{Phys.\ Rev.\ D {\bf 64}, 085015 (2001)}
  %doi:10.1088/1126-6708/2000/02/032
  %[hep-ph/0002221].
  %%CITATION = doi:10.1088/1126-6708/2000/02/032;%%
  %72 citations counted in INSPIRE as of 05 Apr 2020	
  
  
		
		%%%%%%%% String branes and Quantum Gravity %%%%%%
		
\bibitem{Ellis:1992eh} 
  J.~R.~Ellis, N.~E.~Mavromatos and D.~V.~Nanopoulos,
  %``String theory modifies quantum mechanics,''
  \href{https://doi.org/10.1016/0370-2693(92)91478-R}{Phys.\ Lett.\ B {\bf 293}, 37 (1992)}
  %doi:10.1016/0370-2693(92)91478-R
  %[hep-th/9207103].
  %%CITATION = doi:10.1016/0370-2693(92)91478-R;%%
  %250 citations counted in INSPIRE as of 05 Apr 2020	
  
\bibitem{Ellis:1992pm} 
  J.~R.~Ellis, N.~E.~Mavromatos and D.~V.~Nanopoulos,
  %``CPT violation in string modified quantum mechanics and the neutral kaon system,''
  \href{https://doi.org/10.1142/S0217751X96000687}{Int.\ J.\ Mod.\ Phys.\ A {\bf 11}, 1489 (1996)}
  %doi:10.1142/S0217751X96000687
  %[hep-th/9212057].
  %%CITATION = doi:10.1142/S0217751X96000687;%%
  %68 citations counted in INSPIRE as of 05 Apr 2020  	
		
		

\bibitem{Benatti:1998vu}
F.~Benatti and R.~Floreanini,
%``Non-standard Neutral Kaon Dynamics from Infinite Statistics,''
\href{https://doi.org/10.1006/aphy.1998.5896}{Annals Phys.\  \textbf{273}, 58-71 (1999)}
%doi:10.1006/aphy.1998.5896
%[arXiv:hep-th/9811196 [hep-th]].
%27 citations counted in INSPIRE as of 05 Apr 2020		  



\bibitem{Hawking:1982dj}
S.~Hawking,
%``The Unpredictability of Quantum Gravity,''
\href{https://doi.org/10.1007/BF01206031}{Commun.\ Math.\ Phys.\  \textbf{87}, 395-415 (1982)}; \href{https://doi.org/10.1103/PhysRevD.37.904}{Phys.\ Rev.\ D \textbf{37}, 904-910 (1988)}; \href{https://doi.org/10.1103/PhysRevD.53.3099}{Phys.\ Rev.\ D \textbf{53}, 3099-3107 (1996)}; Hawking~S.W. and Hunter~C.J., \href{https://doi.org/10.1103/PhysRevD.59.044025}{Phys.\ Rev.\ D \textbf{59}, 044025 (1999)}



%%%%%%%%%%%%%%%%%%%%%%%%%%%%%%%%%%%%%%%%%%%%%%%%%%%%%%%%%%%%%%%%%%%%%%%%%%%%%%%%%%%%%%%%%

\bibitem{Lisi:2000zt} 
  E.~Lisi, A.~Marrone and D.~Montanino,
  %``Probing possible decoherence effects in atmospheric neutrino oscillations,''
  \href{https://doi.org/10.1103/PhysRevLett.85.1166}{Phys.\ Rev.\ Lett.\  {\bf 85}, 1166 (2000)}
  %doi:10.1103/PhysRevLett.85.1166
  %[hep-ph/0002053].
  %%CITATION = doi:10.1103/PhysRevLett.85.1166;%%
  %243 citations counted in INSPIRE as of 27 Mar 2020

\bibitem{Barenboim:2006xt} 
  G.~Barenboim, N.~E.~Mavromatos, S.~Sarkar and A.~Waldron-Lauda,
  %``Quantum decoherence and neutrino data,''
  \href{https://doi.org/10.1016/j.nuclphysb.2006.09.012}{Nucl.\ Phys.\ B {\bf 758}, 90 (2006)}
  %doi:10.1016/j.nuclphysb.2006.09.012
  %[hep-ph/0603028].
  %%CITATION = doi:10.1016/j.nuclphysb.2006.09.012;%%
  %56 citations counted in INSPIRE as of 27 Mar 2020

\bibitem{Bakhti:2015dca} 
  P.~Bakhti, Y.~Farzan and T.~Schwetz,
  %``Revisiting the quantum decoherence scenario as an explanation for the LSND anomaly,''
  \href{https://doi.org/10.1007/JHEP05(2015)007}{JHEP {\bf 1505}, 007 (2015)}
  %doi:10.1007/JHEP05(2015)007
  %[arXiv:1503.05374 [hep-ph]].
  %%CITATION = doi:10.1007/JHEP05(2015)007;%%
  %14 citations counted in INSPIRE as of 27 Mar 2020


\bibitem{Carpio:2017nui} 
  J.~A.~Carpio, E.~Massoni and A.~M.~Gago,
  %``Revisiting quantum decoherence for neutrino oscillations in matter with constant density,''
  \href{https://doi.org/10.1103/PhysRevD.97.115017}{Phys.\ Rev.\ D {\bf 97}, no. 11, 115017 (2018)}
  %doi:10.1103/PhysRevD.97.115017
  %[arXiv:1711.03680 [hep-ph]].
  %%CITATION = doi:10.1103/PhysRevD.97.115017;%%
  %14 citations counted in INSPIRE as of 27 Mar 2020

\bibitem{Carpio:2018gum} 
  J.~A.~Carpio, E.~Massoni and A.~M.~Gago,
  %``Testing quantum decoherence at DUNE,''
  \href{https://doi.org/10.1103/PhysRevD.100.015035}{Phys.\ Rev.\ D {\bf 100}, no. 1, 015035 (2019)}
  %doi:10.1103/PhysRevD.100.015035
  %[arXiv:1811.07923 [hep-ph]].
  %%CITATION = doi:10.1103/PhysRevD.100.015035;%%
  %2 citations counted in INSPIRE as of 04 Apr 2020		
		


\bibitem{Gomes:2020muc} 
  A.~L.~G.~Gomes, R.~A.~Gomes and O.~L.~G.~Peres,
  %``Quantum decoherence and relaxation in neutrinos using long-baseline data,''
  \href{https://arxiv.org/pdf/2001.09250.pdf}{arXiv:2001.09250 [hep-ph].}
  %%CITATION = ARXIV:2001.09250;%%
  %1 citations counted in INSPIRE as of 27 Mar 2020



%%%%%%%%%%%%%%%%%%%%%%%%%%%%%%%%%%%%%%%%%%%%%%%%%%%%%%%%%%%%%%%%%%%%%%%%%%%%%%%%%%


\bibitem{deOliveira:2013dia} 
  R.~L.~N.~ Oliveira, M.~M.~Guzzo and P.~C.~de Holanda,
  %``Quantum Dissipation and $C\!P$ Violation in MINOS,''
  \href{https://doi.org/10.1103/PhysRevD.89.053002}{Phys.\ Rev.\ D {\bf 89}, no. 5, 053002 (2014)}
  %doi:10.1103/PhysRevD.89.053002
  %[arXiv:1401.0033 [hep-ph]].
  %%CITATION = doi:10.1103/PhysRevD.89.053002;%%
  %10 citations counted in INSPIRE as of 05 Apr 2020



\bibitem{Capolupo:2018hrp} 
  A.~Capolupo, S.~M.~Giampaolo and G.~Lambiase,
  %``Decoherence in neutrino oscillations, neutrino nature and CPT violation,''
  \href{https://doi.org/10.1016/j.physletb.2019.03.062}{Phys.\ Lett.\ B {\bf 792}, 298 (2019)}
  %doi:10.1016/j.physletb.2019.03.062
  %[arXiv:1807.07823 [hep-ph]].
  %%CITATION = doi:10.1016/j.physletb.2019.03.062;%%
  %12 citations counted in INSPIRE as of 05 Apr 2020

\bibitem{Carrasco:2018sca} 
  J.~C.~Carrasco, F.~N.~D\'iaz and A.~M.~Gago,
  %``Probing CPT breaking induced by quantum decoherence at DUNE,''
  \href{https://doi.org/10.1103/PhysRevD.99.075022}{Phys.\ Rev.\ D {\bf 99}, no. 7, 075022 (2019)}
  %doi:10.1103/PhysRevD.99.075022
  %[arXiv:1811.04982 [hep-ph]].
  %%CITATION = doi:10.1103/PhysRevD.99.075022;%%
  %3 citations counted in INSPIRE as of 27 Mar 2020		
  



%%%%%%%%%%%%%%%%%%%%%%%%%%%%Experiments %%%%%%%%%%%%%%%%%%%%%%%%%%%%%%%%%%
	
\bibitem{Acciarri:2015uup} 
  R.~Acciarri {\it et al.} [DUNE Collaboration],
  %``Long-Baseline Neutrino Facility (LBNF) and Deep Underground Neutrino Experiment (DUNE) : Conceptual Design Report, Volume 2: The Physics Program for DUNE at LBNF,''
  \href{https://arxiv.org/pdf/1512.06148.pdf}{arXiv:1512.06148 [physics.ins-det].}
  %%CITATION = ARXIV:1512.06148;%%
  %606 citations counted in INSPIRE as of 27 Mar 2020
  
  %\cite{Abe:2018uyc}
\bibitem{Abe:2018uyc}
K.~Abe \textit{et al.} [Hyper-Kamiokande],
%``Hyper-Kamiokande Design Report,''
\href{https://arxiv.org/abs/1805.04163}{arXiv:1805.04163 [physics.ins-det].}
%193 citations counted in INSPIRE as of 19 Jun 2020


%%%%%%%%%%%%%%%%%%%%%%%%%%%%%%%%%%%%%%%%%%%%%%%%%%%%%%%%%%%%%%%%%5



%
\bibitem{Gago:2002na}
A.~M.~Gago, E.~M.~Santos, W.~J.~C.~Teves and R.~Zukanovich Funchal,
%``A Study on quantum decoherence phenomena with three generations of neutrinos,''
\href{https://arxiv.org/abs/hep-ph/0208166}{arXiv:hep-ph/0208166 [hep-ph]}.
%42 citations counted in INSPIRE as of 31 Oct 2020
%LaTeX (EU) 




%%%%%%%%%%%%%%%%%%%%%%%%%%%%%%%%%%%%%%%%%%%%%%%%%%%%%%%%%%%%%%%%%%%%%%%%%%%%%%5

\bibitem{Maki:1962mu}
Z.~Maki, M.~Nakagawa and S.~Sakata,
%``Remarks on the unified model of elementary particles,''
\href{https://academic.oup.com/ptp/article/28/5/870/1858382}{Prog. Theor. Phys. \textbf{28} (1962), 870-880}
%doi:10.1143/PTP.28.870
%4345 citations counted in INSPIRE as of 04 Jul 2021




%%%%%%%%%%%%%%%%%%%%%%%%%%%%555%% Nufit %%%%%%%%%%%%%%%%%%%%%%%%%%%%%%%%%%%%%%%%%%
		\bibitem{Nufit}
		\href{http://www.nu-fit.org/}{http://www.nu-fit.org}
		
		
		
		
		
		
%%%%%%%%%%%%%%%%%%%%%%%%%%%%%%%%%%%%%%%%%%%%%%%%%%%%%%%%%%%%%%%%%
%%%%%%%%%%%%%%%%%%%%%%  GLoBES %%%%%%%%%%%%%%%%%%%%%%%%%%%%%%%%


	\bibitem{Huber:2004ka} 
  P.~Huber, M.~Lindner and W.~Winter,
  %``Simulation of long-baseline neutrino oscillation experiments with GLoBES (General Long Baseline Experiment Simulator),''
  \href{https://doi.org/10.1016/j.cpc.2005.01.003}{Comput.\ Phys.\ Commun.\  {\bf 167}, 195 (2005)}
  %doi:10.1016/j.cpc.2005.01.003
  %[hep-ph/0407333].
  %%CITATION = doi:10.1016/j.cpc.2005.01.003;%%
  %568 citations counted in INSPIRE as of 27 Mar 2020
	
	
\bibitem{Huber:2007ji} 
  P.~Huber, J.~Kopp, M.~Lindner, M.~Rolinec and W.~Winter,
  %``New features in the simulation of neutrino oscillation experiments with GLoBES 3.0: General Long Baseline Experiment Simulator,''
  \href{https://doi.org/10.1016/j.cpc.2007.05.004}{Comput.\ Phys.\ Commun.\  {\bf 177}, 432 (2007)}				
		
		

%%%%%%%%%%%%%%%%%%%%%%%%%%%%%%%% nuSQuIDS %%%%%%%%%%%%%%%%%%%%%%%%%%%%%%%%%

\bibitem{Delgado:2014kpa} 
  C.~A.~Arg\"uelles Delgado, J.~Salvado and C.~N.~Weaver,
  %``A Simple Quantum Integro-Differential Solver (SQuIDS),''
  \href{https://doi.org/10.1016/j.cpc.2015.06.022}{Comput.\ Phys.\ Commun.\  {\bf 196}, 569 (2015)}
  %doi:10.1016/j.cpc.2015.06.022
  %[arXiv:1412.3832 [hep-ph]].
  %%CITATION = doi:10.1016/j.cpc.2015.06.022;%%
  %35 citations counted in INSPIRE as of 05 Apr 2020

%%%%%%%%%%%%%%%%%%%%%%%%%%%%%%%%%%%%%%%%%%%%%%%%%%%%%%%%%%%%%%%%%%%%%%%%%%%%%%
		



%%%%%%%%%%%%%%%%%%%%%%%%%%%%%%% DUNE %%%%%%%%%%%%%%%%%%%%%%%%%%%%%%%%%%%%%%%%%%%%%%%%%

\bibitem{Alion:2016uaj} 
  T.~Alion {\it et al.} [DUNE Collaboration],
  %``Experiment Simulation Configurations Used in DUNE CDR,''
  \href{https://arxiv.org/pdf/1606.09550.pdf}{arXiv:1606.09550 [physics.ins-det]}.
  %%CITATION = ARXIV:1606.09550;%%
  %69 citations counted in INSPIRE as of 27 Mar 2020




%%%%%%%%%%%%%%%%%%%%%%%%%%%%%%%%% T2HK %%%%%%%%%%%%%%%%%%%%%%%%%%%%%%%%%%

\bibitem{Huber:2007em}
P.~Huber, M.~Mezzetto and T.~Schwetz,
%``On the impact of systematical uncertainties for the CP violation measurement in superbeam experiments,''
\href{https://iopscience.iop.org/article/10.1088/1126-6708/2008/03/021}{JHEP \textbf{03}, 021 (2008)}
%doi:10.1088/1126-6708/2008/03/021
%[arXiv:0711.2950 [hep-ph]].
%61 citations counted in INSPIRE as of 18 Sep 2020
		
		
		
%%%%%%%%%%%%%%%%%%%%%%%%%%T2K%%%%%%%%%%%%%%%%%%%%%%%%%%%%%%%%%%

		
\bibitem{Abe:2019vii}
K.~Abe \textit{et al.} [T2K],
%``Constraint on the matter–antimatter symmetry-violating phase in neutrino oscillations,''
\href{https://doi.org/10.1038/s41586-020-2177-0}{Nature \textbf{580}, no.7803, 339-344 (2020)}
%doi:10.1038/s41586-020-2177-0
%[arXiv:1910.03887 [hep-ex]].
%20 citations counted in INSPIRE as of 01 May 2020


		
		
%%%%%%%%%%%%%%%%%%%%%%%%%%%% NOvA %%%%%%%%%%%%%%%%%%%%%%%%%

\bibitem{novaresults}
A. Himmel, \href{https://indico.fnal.gov/event/43209/contributions/187840/attachments/130740/159597/NOvA-Oscilations-NEUTRINO2020.pdf}{“New Oscillation Results from the NOvA Experiment -- presented at Neutrino2020,” (2020).}






  




\bibitem{Diaz:2020aax} 
  F.~N.~D\'iaz, J.~Hoefken and A.~M.~Gago,
  %``Effects of the Violation of the Equivalence Principle at DUNE,''
  \href{https://journals.aps.org/prd/abstract/10.1103/PhysRevD.102.055020}{Phys. Rev. D \textbf{102} (2020) no.5, 055020}
  %%CITATION = ARXIV:2003.13712;%%


		
		
%
\bibitem{Coloma:2018idr}
P.~Coloma, J.~Lopez-Pavon, I.~Martinez-Soler and H.~Nunokawa,
%``Decoherence in Neutrino Propagation Through Matter, and Bounds from IceCube/DeepCore,''
\href{https://link.springer.com/article/10.1140%2Fepjc%2Fs10052-018-6092-6}{Eur. Phys. J. C \textbf{78} (2018) no.8, 614}
%doi:10.1140/epjc/s10052-018-6092-6
[arXiv:1803.04438 [hep-ph]].
%25 citations counted in INSPIRE as of 09 Jun 2021		
		
		
		
%%%%%%%%%%%%%%%%%%%%%%%%%%%%%%%%%%%%%%%%%%%%%%%%%%%%%%%%%%%%%%


\bibitem{Rodejohann:2011vc}
W.~Rodejohann and J.~W.~F.~Valle,
%``Symmetrical Parametrizations of the Lepton Mixing Matrix,''
\href{https://journals.aps.org/prd/abstract/10.1103/PhysRevD.84.073011}{Phys. Rev. D \textbf{84} (2011), 073011}
%doi:10.1103/PhysRevD.84.073011
%[arXiv:1108.3484 [hep-ph]].
%72 citations counted in INSPIRE as of 01 Jul 2021



		

%\cite{Tanabashi:2018oca}
\bibitem{Tanabashi:2018oca}
M.~Tanabashi \textit{et al.} [Particle Data Group],
%``Review of Particle Physics,''
\href{https://journals.aps.org/prd/abstract/10.1103/PhysRevD.98.030001}{Phys. Rev. D \textbf{98}, no.3, 030001 (2018)}
%doi:10.1103/PhysRevD.98.030001
%6911 citations counted in INSPIRE as of 09 Jun 2021
%\cite{Rodejohann:2011vc}





 
  



		

		
		
%%%%%%%%%%%%%%%%%%%%%%%%%%%%%%%%%%%%%%%%%%%%%%%%%%%%%%%%%%%%%%%%%%%%%%%%%%%%%%%%%%%
		




	\end{thebibliography}
\end{document}